\RequirePackage{fix-cm}
\documentclass[smallextended]{svjour3}       
\smartqed  
\usepackage{graphicx}
\title{Insert your title here
}

\author{Lili Wei         \and
        Heqing Huang     \and
        Shing-Chi Cheung \and
        Kevin Li
}

\institute{Corresponding author - Lili Wei \at
              Department of Electrical and Computer Engineering, McGill University, Canada
              \\
              \email{lili.wei@mcgill.ca}           
           \and
          Corresponding author - Heqing Huang \at
              Department of Computer Science,
              City University of Hong Kong,
              China\\
              \email{heqhuang@cityu.edu.hk}
           \and
           Shing-Chi Cheung \at
              Department of Computer Science and Engineering, The Hong Kong University of Science and Technology, China
              \\
              \email{scc@cse.ust.hk}
              \and
              Kevin Li \at
              Department of Electrical and Computer Engineering, McGill University, Canada
              \\
              \email{kevin.li3@mail.mcgill.ca}
}

\date{Author pre-print copy. The final publication is available at Springer via:\\
\url{https://link.springer.com/article/10.1007/s10664-024-10607-9}}
\usepackage{multirow}
\usepackage{graphicx}
\usepackage{cite}
\usepackage{amsmath,amssymb,amsfonts}
\usepackage{algorithmic}
\usepackage{graphicx}
\usepackage{textcomp}
\usepackage{xcolor}
\usepackage{balance}
\usepackage{listings}
\usepackage{booktabs}
\usepackage{enumitem}
\usepackage[hyphens]{url}
\usepackage[singlelinecheck=false]{caption}
\def\BibTeX{{\rm B\kern-.05em{\sc i\kern-.025em b}\kern-.08em
    T\kern-.1667em\lower.7ex\hbox{E}\kern-.125emX}}
\newcommand{\revision}[1]{{\color{blue} #1}}

\newenvironment{revised}
{
    \color{blue}
}

\begin{document}

\title{How Far are App Secrets from Being Stolen? A Case Study on Android\\
}


\maketitle

\begin{abstract}

Android apps can hold secret strings of themselves such as cloud service credentials or encryption keys.
Leakage of such secret strings can induce unprecedented consequences like monetary losses or leakage of user private information.
In practice, various security issues were reported because many apps failed to protect their secrets.
However, little is known about the types, usages, exploitability, and consequences of app secret leakage issues.
While a large body of literature has been devoted to studying \textit{user private information leakage}, there is no systematic study characterizing \textit{app secret leakage issues}.
How far are Android app secrets from being stolen?

To bridge this gap, we conducted the first systematic study to characterize app secret leakage issues in Android apps based on 575 potential app secrets sampled from 14,665 popular Android apps on Google Play.
We summarized the common categories of leaked app secrets, assessed their security impacts and disclosed app bad practices in storing app secrets.
We devised a text mining strategy using regular expressions and demonstrated that numerous app secrets can be easily stolen, even from the highly popular Android apps on Google.
In a follow-up study, we harvested 3,711 distinct exploitable app secrets through automatic analysis.
Our findings highlight the prevalence of this problem and call for greater attention to app secret protection.

\end{abstract}

\section{Introduction}
In the past decade, we witnessed the rapid growth of Android app markets.
Along with their popularity, apps become more complex~\cite{app_complicated}.
They are integrated with diverse functionalities that rely on increasingly complicated techniques. 
The growing complexity brings new security challenges in Android app development.
Existing studies focus mostly on detecting and alleviating \textbf{user private information leakage} in Android apps such as detecting information leaks by tracking information flows~\cite{arzt2014flowdroid,enck2014taintdroid,li2015iccta,gamba2020analysis}, and assessing the security concerns of the Android permission models~\cite{tuncay2018resolving,felt2011android,shen2021can,wijesekera2015android,reardon201950}.
Compared to the large amount of literature that has been devoted to user private information leakage, there is a lack of understanding on \textbf{app secret leakage}.

Android apps can hold a variety of secret strings.
Such strings represent sensitive information belonging to the host apps.
For example, an emerging number of apps integrate third-party cloud services~\cite{zuo2019does}.
Such an app needs to use its credential to invoke the APIs of the subscribed cloud services. The credential is an app secret representing app identities.

The leakage of app secrets can cause unprecedented consequences.
In practice, many apps failed to properly protect their secrets.
Various researchers and hackers reported cases where they successfully extracted app secrets~\cite{google_maps_key_blog,key_leak_blog,hackerone1,FCM_takeover,zuo2019does}.
For example, a hacker was able to send arbitrary notification messages to billions of users by harvesting Firebase Cloud Messaging service keys in Android apps~\cite{FCM_takeover}.
Zuo et al.~\cite{zuo2019does} also reported that including root keys of AWS S3 Bucket in Android apps would allow attackers to manipulate apps' sensitive resources that were hosted on AWS servers.
Several papers identified cases where app secrets were stored as plain texts in Android apps~\cite{chen2014oauth,al2019oauthlint}.
These papers either reported a limited number of secret leakage cases or focused on issues specific to certain cloud services.
None of them systematically study the app secret leakage issues in Android apps.

In this paper, we present the first empirical analysis on the problem of app secret leakage.
We aim to summarize the diverse types of leaked app secrets as well as demonstrating their prevalence.
There can be different ways to steal app secrets, such as dynamic analysis or network traffic spoofing. In this paper, we explore stealing app secrets from APK files via static analysis due to its simplicity and efficiency.
We find that thousands of app secrets can be stolen from popular apps on Google Play~\cite{googleplay} using simple static analysis techniques.

Our study aims to characterize app secret leakage issues by answering the following research questions:

\begin{itemize}[leftmargin=*]
    \item \textbf{RQ1:} What types of secrets can be stolen from the Android apps? How are they used in Android apps?
    \item \textbf{RQ2:} How can the leaked secrets be exploited? What are the consequences of app secret leakage issues?
    \item \textbf{RQ3:} What are the bad practices in storing the secrets?
\end{itemize}
The answers to RQ1 and RQ2 provide a systematic overview of app secret leakage issues in Android by analyzing what types of app secrets can be leaked and assessing their security impact.
The study of RQ3 enables us to identify bad practices in storing app secrets so that developers can avoid such practices or take countermeasures.

To answer these research questions, we conducted an empirical study based on the manual inspection of 575 app secret candidates sampled from 14,665 popular Android apps on Google Play.
We analyzed how these secret candidates were used in the Android apps and collected relevant online discussions.
Our study reveals that many Android apps contain app secrets that can be easily extracted by simple text mining approaches such as regular expression matching.
In our study, we used regular expressions and identified a variety of app secrets including app credentials of third-party or app-private cloud services, and static encryption keys.
Leaking these secrets can induce various security consequences such as monetary losses, denial of service, or user private information leakage.
We find that even Android apps with millions of downloads on Google Play often fail to protect their secrets properly.
They hard code app secrets in their apps without protection and tend to put a collection of app secrets in the same file.
The lack of proper protection makes it easy for attackers to extract app secrets from victim Android apps.

We further conducted a study to detect exploitable app secrets for commonly-used third-party cloud services in Android apps to assess the prevalence of exploitable app secrets in the wild.
In total, we identified 3,711 distinct exploitable app secrets in the latest version of 23,041 most  popular apps on Google Play.
Our analysis of the exploitable secrets revealed interesting findings:
One of our identified exploitable secrets was used in as many as 53 apps.
Once the secret is comprised and abused, all the apps will be affected.
We also discovered evidences for frauds of app secrets even between the most popular apps.
For example, we identified a leaked app secret in a social app with over 1 billion downloads, which was also used in another six apps with millions of downloads developed by different companies.
With our report, the social app quickly confirmed and fixed the issue.

Our paper presents the first empirical study demonstrating that app secrets are not far from being stolen.
While there exist various empirical analysis on the problem of leaked secrets in the context of GitHub projects~\cite{meli2019bad}, our study is the first empirical analysis of the problem in the context of Android apps.
Our study performs in-depth analyses on app secret leakage issues in the wild and provides a first systematic overview of the issues.
Although hard-coded secrets are a long-standing security problem, our study results show that the problem is still prevalent in Android apps and developers made limited efforts to protect them.
Our findings pave the way for future studies in this topic.
To summarize, we make the following contributions in this paper:
\begin{itemize}[leftmargin=*]
    \item We conducted the first systematic study to characterize the leakage of app secrets in Android apps.
    We categorized the leaked app secrets according to their usages, assessed their security impacts, and summarized developers' bad practices in storing app secrets.
    \item  We constructed regular expressions for app secret mining and showed that they can successfully detect many app secrets kept by popular apps. We presented possible measures that can be taken to protect app secrets.
    \item We found that many app secrets detected in highly popular Android apps on Google Play are exploitable. We reported app secret leakage issues and discussed the strategies adopted by the app companies.
\end{itemize}
\section{Background}
\subsection{User Private Information v.s. App Secrets}
User private information refers to sensitive data owned by Android users, such as contacts, photos, and calendars. This information is typically stolen at runtime by malicious apps that collect and transmit it to their servers.
Various techniques have been proposed to detect user private information leakage in Android apps~\cite{li2015iccta,arzt2014flowdroid,enck2014taintdroid,felt2011android,wijesekera2015android,shen2021can,gamba2020analysis,tuncay2018resolving,reardon201950}.
Android also provides permission management systems to protect user private information: an app can only access certain user information after receiving users' consents.

In comparison, app secrets are sensitive information owned by Android apps.
In this paper, we focus on the leakage issues of string literal app secrets.
They either represent the identity of Android apps used for authentication or serve as master keys to protect other app-private information.
Typical examples of app secrets include app credentials of cloud services and encryption keys.
We study the app secrets that are stored in Android app code.
The leakage of some app secrets can also induce user private information leakage.
For example, the app credentials of a cloud service may allow an attacker to access the victim app's user information that is hosted on the cloud service.
However, app secrets themselves do not represent any information of their app users.

\subsection{Reverse Engineering for Android Apps}
Android apps are compiled into APK files and distributed to app stores such as Google Play~\cite{apk}.
An APK file is an archive that contains the app's resources and code.
Android programs are eventually compiled into Dalvik bytecode in .dex files which are executable files of Android Runtime.
Various tools are available to reverse-engineer APK files such as ApkTool~\cite{apktool} and Bytecode Viewer~\cite{bytecodeviewer}.
ApkTool can unpack APK files by disassembling the resources and .dex files.
ApkTool decompiles .dex files into smali code, a human-readable representation of Dalvik bytecode~\cite{smali}.
Bytecode Viewer decompiles .dex files into Java source code and provides a GUI code viewer.
In our study, we used both of these tools.
We used ApkTool to unpack our subject apps and built analyzers upon the unpacked APKs. We used Bytecode Viewer for the manual inspection of app code.



\section{Characteristic Study Setup}
\subsection{Data Collection}\label{ssec:data_collection}

To characterize the leaked secrets in Android apps, we collected a set of leaked secrets in Android apps.
Techniques proposed by existing work are restricted to several predefined types of secrets~\cite{zuo2019does,meli2019bad}.
Since our study aims to understand what kinds of app secrets can be leaked in Android apps, it is inappropriate to restrict the secrets to specific types.
Therefore, we need an alternative approach that can obtain diverse kinds of secrets.
To achieve this goal, we devised a search-based approach to first collecting potential app secrets (secret candidates) in reverse-engineered Android apps, and then identifying app secrets from the candidates. 
Our approach can detect more kinds of app secrets compared with existing techniques.
Specifically, we obtained the secrets in three steps.

\textbf{Step 1: Collecting Secret Candidates.}
In this step, we collected strings in Android apps that are potential app secrets (i.e., secret candidates).
To achieve this, we collected four Android app datasets consisting of the most popular free Android apps of each category on Google Play in 2017, 2018, 2019, and 2020.
For the datasets of 2017 and 2018, we used app datasets in a previous study~\cite{wei2019pivot}.
For the datasets of 2019 and 2020, we crawled the top 100 most popular apps in each category by ourselves.
During this process, we ignored those apps that cannot be downloaded due to server errors or regional restrictions.
In total, we collected 14,665 Android apps.

\begin{figure*}
	\centering
	\includegraphics[width=\linewidth]{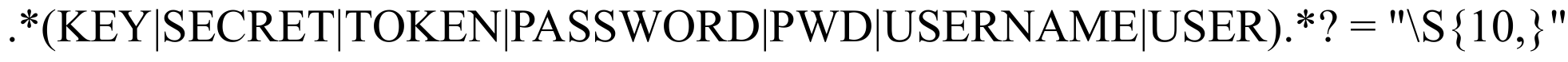}
	\captionof{lstlisting}{\centering The Regular Expression to Extract Secret Candidates}
	\label{listing:regex}
\end{figure*}

We leveraged ApkTool to reverse-engineer all the collected apps~\cite{apktool}.
We then searched for secret candidates with regular expressions, 
following a similar approach adopted by Meli et al.~\cite{meli2019bad}.
However, our search strategy is different. 
To characterize secret leakage in git services, Meli et al. ~\cite{meli2019bad} adopted a set of regular expressions matching credentials with strong patterns concerning a limited number of cloud services.
In contrast, our study aims to understand what kinds of app secrets can be leaked and thus cannot restrict the secret types. 
To this end, we designed a regular expression as given in Listing~\ref{listing:regex}.
Intuitively, this regular expression searches for definitions of variables whose names contain sensitive keywords such as KEY or SECRET, and the variable is defined by a string literal of length over ten.
We chose the keywords according to the names of program secret and credential information~\cite{GUO2021102156} and empirically set the length of the string literal to over ten since we found that a huge amount of noises would be included if the string lengths are not restricted.
Our work can identify more types of app secrets compared with the approach adopted by Meli et al.~\cite{meli2019bad}.

\begin{figure*}
	\centering
	\includegraphics[width=\linewidth]{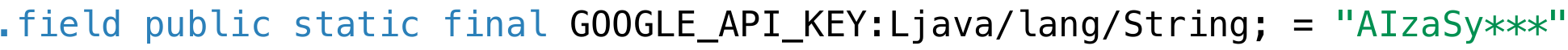}
	\captionof{lstlisting}{An Example Line of Code That Matches the Regular Expression in Listing~\ref{listing:regex}. This line was extracted from an app with over 100 million downloads on Google Play.}
	\label{listing:regex_example}
\end{figure*}

The string literals to define the variables are our extracted secret candidate.
For example, the code in Listing~\ref{listing:regex_example} shows a line of code that matches the regular expression.
It was adapted from a real line that was extracted in our characteristic study.
It is a line of smali code~\cite{smali} that defines a static field named \texttt{GOOGLE\_API\_KEY} by a string literal ``\texttt{AIzaSy***}''.
The string literal ``\texttt{AIzaSy***}'' is the secret candidate.
We present part of the string literal as \texttt{***} in this paper to avoid leaking the sensitive information of the app.

This regular expression can induce both false positives and false negatives. 
There are lines defining non-secret variables that match the regular expression.
There are also secrets that are not defined with the keywords.
First, apps can be obfuscated.
Obfuscated variable names may not contain any of the keywords.
Second, secrets can be obtained by string manipulations.
Despite the limitation of using such a regular expression to search for app secrets, it is still useful to collect secret candidates that are used for further manual inspection.



\textbf{Step 2: Ranking Secret Candidates.}
In the first step, we obtained 1,627,053 lines of code that matched the regular expression.
However, many of these extracted lines did not contain app secrets.
We checked the first 100 of the extracted lines, among which only two contain app secrets (2\%).
For the other candidates, they are mostly keys in key-value pairs or public URLs.
Such immense noises can induce great overhead in our manual inspection.
This motivates us to filter the whole dataset to obtain a subset for our manual inspection.

It is non-trivial to automatically distinguish app secrets from a large number of noises since app secrets can be of different forms.
There is no golden rule that can precisely identify all the app secrets from the secret candidates.
To resolve this problem, we prioritized the secret candidates based on the following intuition:
A secret candidate is more likely to be an app secret if (1) the characters in it are more diverse, and (2) it contains fewer English words.
Given a secret candidate $s$, its ranking score $R(s)$ can be computed as:
$$R(s) = R_\mathit{diversity}(s) + R_\mathit{words}(s)$$
where $R_\mathit{diversity}(s)$ captures the diversity of the characters in $s$.
It is computed as the standard deviation of the ASCII values of all the characters in $s$.
$R_\mathit{words}(s)$ captures intuition (2).
To compute $R_\mathit{words}(s)$, we first split $s$ by non-letter symbols and then split the resulted substrings according to camel case.
Then we matched each substring with a English dictionary to check whether the substring is an English word.
$R_\mathit{words}(s)$ is computed as the proportion of non-word substrings among all the substrings.
Higher $R_\mathit{words}(s)$ means that the proportion of words is lower in $s$.

Note that our ranking strategy is not perfect.
There can be app secrets that do not have high ranking scores as well as non-secrets having high ranking scores.
However, it is acceptable to have an imprecise ranking strategy since we will further inspect the candidate secrets.
In the following section, we discuss our secret candidate sampling approach based on the ranking scores.
The ranking scores are used to increase the probability to sample secret candidates that are more likely to be app secrets.
In Section~\ref{ssec:RQ1}, we show that, with the help of our ranking strategy, 56.9\% (327/575) of our sampled secret candidates were identified as app secrets, which is much larger than that of pure random sampled datasets (2\%, 2/100).

    The ranking strategy still results in a low precision but it leads to a broader range of collected secrets for our later manual analysis.
    The regular expression is designed to include as many different types of secrets as possible for our empirical study.
    This introduced a significant amount of noises, making it impossible to manually analyze.
    To mitigate this issue, we proposed the ranking strategy.
    The ranking strategy is a pre-processing step before our manual analysis.
    Since the precision of the approach is not the focus of our study, we did not further improve the precision of our approach.
    \revision{Although this pre-processing step is not sound, we further conducted a rigorous open coding process to ensure the reliability of our empirical findings (Section~\ref{ssec:dataanalysis}).}

    There exist more precise approaches.
    For example, Zuo et al.~\cite{zuo2019does} proposed a static analysis tool, LeakScope, to extract leaked secrets from apps that are input to specific authentication APIs.
    \revision{LeakScope is the only available leaked secret detection tool that specifically focuses on Android.}
    However, LeakScope only work for specific types of secrets:
    it can only detect leaked secrets of Azure Storage, Azure Notification Hub and Amazon AWS S3.  
    Using it will limit the scope of our empirical study.
    \begin{revised}
        In addition, there exist some secret detection techniques for open-source repositories on GitHub~\cite{meli2019bad,passfinder}. Meli et al.~\cite{meli2019bad} leveraged regular expressions to identify leaked secrets, which is similar to our approach in Section~\ref{sec:detection}. Feng et al.~\cite{passfinder} proposed PassFinder, a tool that uses machine learning techniques to identify leaked secrets. However, PassFinder is trained on source code, whereas our study focuses on closed-source Android apps that are typically obfuscated. Consequently, PassFinder is not directly applicable to our research.
    \end{revised}

\textbf{Step 3: Sampling Secret Candidates.}
It is impractical to manually inspect all of the extracted secret candidates that are extracted in Step 1.
As a result, we sampled a subset of the secret candidates to form our characteristic study dataset.
We sampled the secret candidates according to their ranking scores.
Specifically, we used the computed ranking scores as weights and performed a weighted random sampling over all the secret candidates.
We sampled 300 secret candidates for further inspection.
We also included all the secret candidates that contain only numbers (275 strings) for further inspection since strings with only numbers are likely app secrets but may have low ranking scores.
It is also affordable for us to manually inspect 275 secret candidates.
In total, we obtained 575 secret candidates as our characteristic study dataset.
This sample size has a confidence level of 95\% with a 4.09\% margin of error for an unlimited-large population~\cite{samplesize}.
This exceeds the commonly adopted confidence level and margin of error of the sample sizes adopted by other empirical studies in the computer science discipline~\cite{harty2021logging,liu2021characterizing}.

\subsection{Data Analysis}\label{ssec:dataanalysis}
To inspect the characteristic study dataset, we followed the process of open coding~\cite{holton2007coding}.
Two of our co-authors independently checked each of the sampled lines and answered the research questions.
For each of the secret candidates, the co-authors (1) referred back to the location where the candidate was extracted, (2) analyzed the surrounding code and the dataflow of the concerned variables, and (3) searched for related online resources to answer the three research questions.
When both co-authors checked every 100 secret candidates, they cross-checked their categorization results, resolved disagreements and updated the code book.
In the end, the co-authors reached a consensus on all the inspected candidate secrets.
In the following, we present our answers to each of the research questions based on our manual inspection.
\section{Results}
\subsection{RQ1 Secret Types}\label{ssec:RQ1}

\begin{table*}[t]
\centering
\caption{\centering Categories of Secret Candidates According to Their Usages}
\label{tab:RQ1Category}
\resizebox{\textwidth}{!}{%
\begin{tabular}{|l|l|l|r|l|}
\hline
\multicolumn{3}{|l|}{\textbf{Category}}                                                                                               & \multicolumn{1}{l|}{\textbf{\#Candidates}} & \textbf{Example} \\ \hline
\multirow{4}{*}{App credentials for third-party cloud services} & \multirow{2}{*}{App credentials for Authentication} & Single-factor & 32                                   & Listing~\ref{listing:regex_example}\\ \cline{3-5} 
                                                                &                                                     & Multi-factor  & 98                                  & Listing~\ref{listing:twitter_keys}, \ref{listing:cloudinary_keys}\\ \cline{2-5} 
                                                                & \multicolumn{2}{l|}{Public app keys for cloud services}             & 127                                  &Listing~\ref{listing:flurry_keys}\\ \cline{2-5} 
                                                                & \multicolumn{2}{l|}{\textbf{Total}}                                          & \textbf{257}                                  & -                 \\ \hline
\multicolumn{3}{|l|}{Encryption keys}                                                                                                 & 25                                   &Listing~\ref{listing:encryption_example}\\ \hline
\multicolumn{3}{|l|}{App credentials for app-private back-end services}                                                              & 24                                   &Listing~\ref{listing:app_apecific_code}\\ \hline
\multicolumn{3}{|l|}{App credentials for cloud services that are not maintained}                                                      & 21                                   & -                 \\ \hline
\multicolumn{3}{|l|}{Obviously not app secrets}                                                                                       & 188                                  & -                 \\ \hline
\multicolumn{3}{|l|}{Others}                                                                                                          & 60                                   & -                 \\ \hline
\multicolumn{3}{|l|}{\textbf{Total}}                                                                                                           & \textbf{575}                                     & -                 \\ \hline
\end{tabular}%
}
\end{table*}

In this research question, we aim to categorize the secret candidates according to their usages.
Understanding the usages of the secrets is crucial for determining whether the leaked secrets are sensitive and susceptible to be exploited.
We leveraged ApkTool to reverse-engineer the apps, manually checked the data flow regarding each potential secret, and examined their usages.
For those secrets related to third-party libraries, we also searched for their documents to better understand their usages.
Table~\ref{tab:RQ1Category} shows the taxonomy that we built based on the characteristic study dataset.
We introduce each of the categories below.

\textbf{App credentials for third-party cloud services.}
The majority (257 out of 575) of the inspected secrets are related to third-party cloud services.
We categorized a candidate into this category if (a) it is passed to authentication-related parameters of APIs in SDKs of third-party cloud services, or (b) it is used as part of the authentication header in requests to cloud service APIs.
Among them, 130 candidates are app credentials for cloud service authentication and the other 127 are public app keys.
We list all the concerned cloud services of these credentials in Appendix.

\begin{figure}
	\centering
	\includegraphics[width=\linewidth]{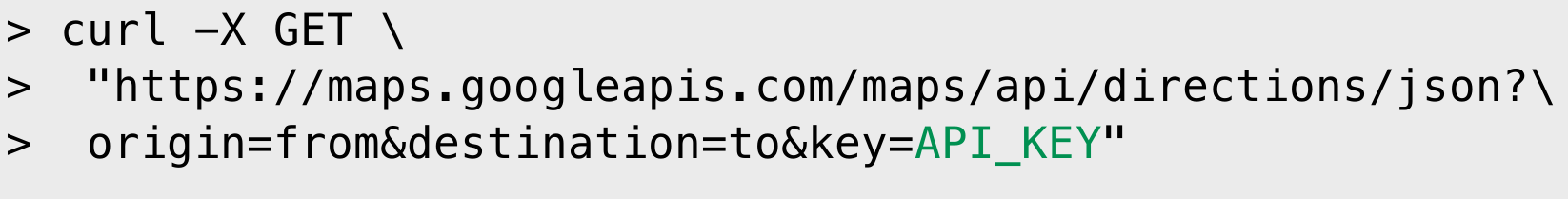}
	\captionof{lstlisting}{The Command to Invoke Directions API of Google Maps Platform Where \textit{API\_KEY} Should Be Replaced with a Valid Google Maps API Key.}
	\label{listing:google_maps_command}
\end{figure}
\begin{figure}
	\centering
	\includegraphics[width=\linewidth]{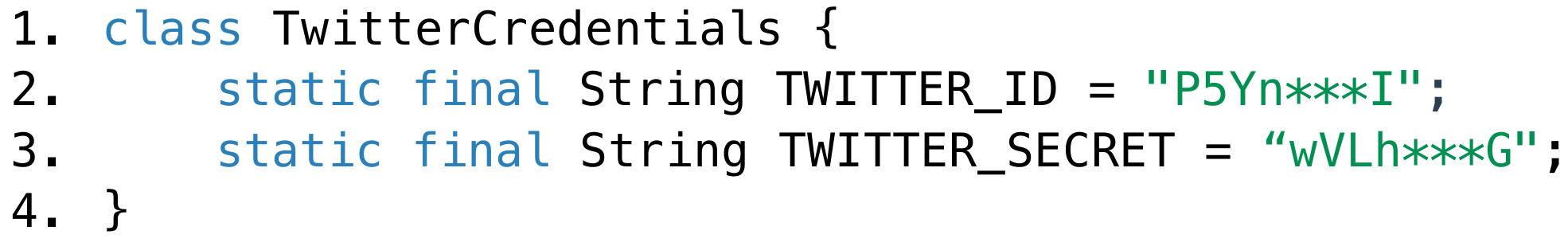}
	\captionof{lstlisting}{Anonymized Twitter Client ID and Client Secret in Our Dataset.\protect\footnotemark The secrets were extracted from an app with over 10 million downloads on Google Play.}
	\label{listing:twitter_keys}
\end{figure}

\textit{App credentials for cloud service authentication.}
App credentials for cloud service authentication can be single-factor or multi-factor.
Single-factor credentials alone can authenticate the corresponding cloud API invocations.
The example shown in Listing~\ref{listing:regex_example} is a single-factor app credential for Google Maps Platform API.
This secret can be used to invoke the Geolocation API of Google Maps Platform with the command shown in Listing~\ref{listing:google_maps_command}.

\begin{figure}
	\centering
	\includegraphics[width=\linewidth]{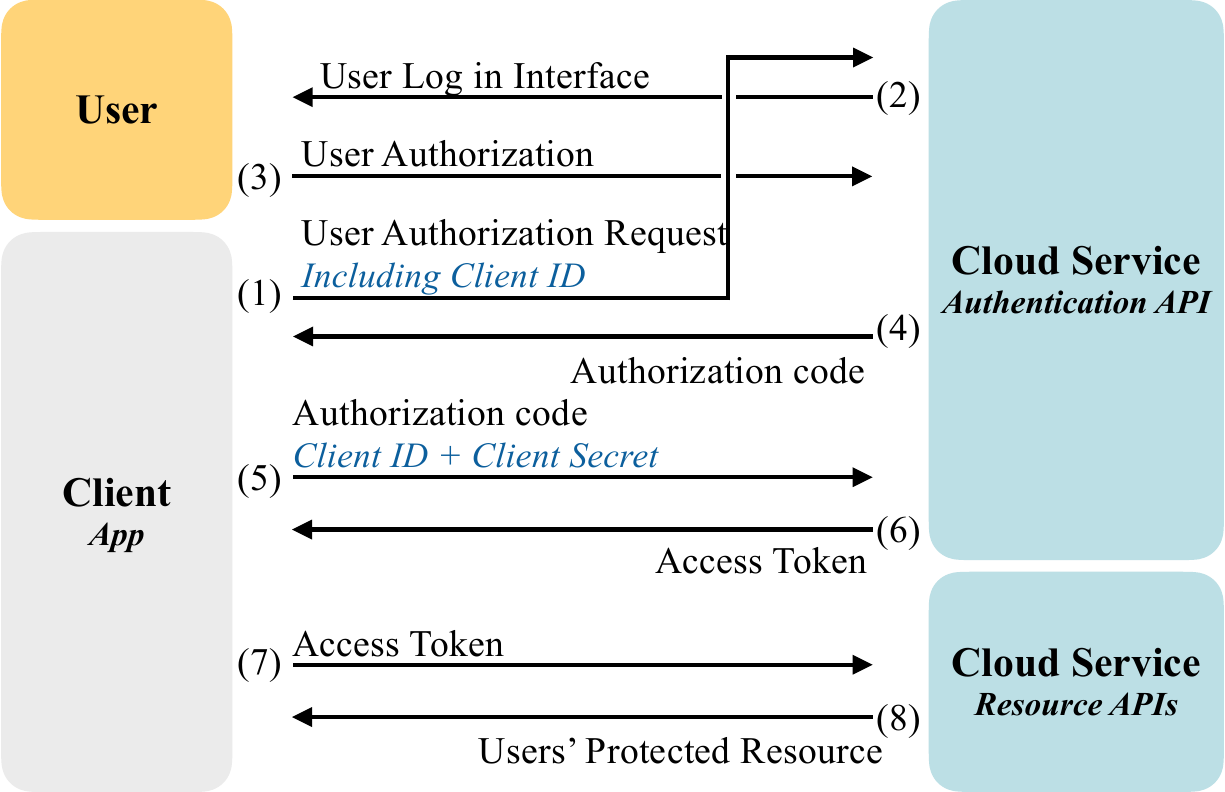}
	\caption{Workflow of Authorization Code Grant Type for OAuth.}
	\label{fig:oauth}
\end{figure}

\begin{figure}[t]
	\centering
	\includegraphics[width=\linewidth]{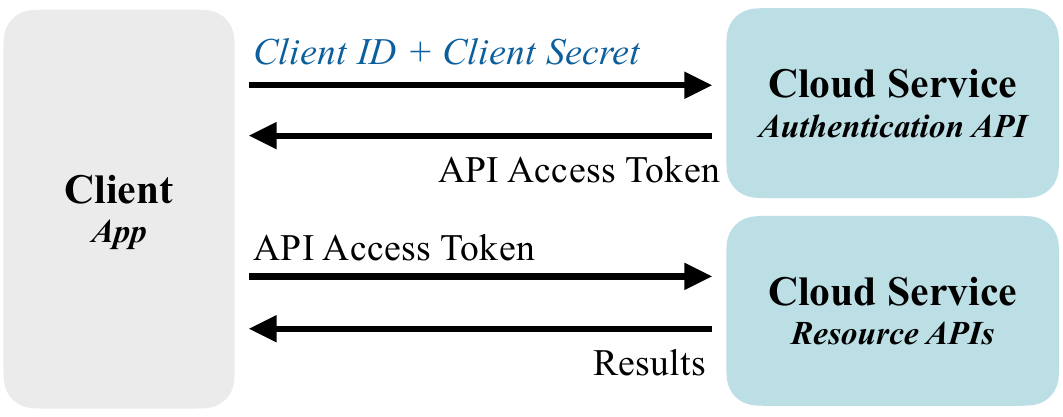}
	\caption{Workflow of Client Credentials Grant Type for OAuth.}
	\label{fig:token_workflow}
\end{figure}

There are also multi-factor app credentials for cloud service authentication.
In this scenario, Multiple credentials are used together to authenticate cloud API invocations.
We found 98 multi-factor app credentials in our study dataset.
Multi-factor app credentials can be used in different ways.
One common use for multi-factor app credentials is OAuth, which is an authorization protocol that allows users to share their information on cloud services with third-party clients such as Android apps without giving their passwords~\cite{oauth}.\footnote{In this paper, we focus on OAuth 2.0 since it is the latest OAuth standard and is widely used by different cloud services.}
There are different workflows for OAuth.
Figures~\ref{fig:oauth} and \ref{fig:token_workflow} give two commonly used OAuth workflows.
Figure~\ref{fig:oauth} shows the workflow of the authorization code grant type for OAuth~\cite{oauth_code_type}.
This workflow is recommended to be used for mobile applications. It is adopted by many companies like Google, Facebook and Twitter.
The authorization process involves three parties: a user, a client (an app), and a cloud service.
In the OAuth process, an app uses both a client ID and a client secret to verify its identity~\cite{oauth_client_credentials}.
We categorized client IDs and client secrets into this category because both of them are needed together for cloud service authentication.
However, we found many leaks of  both client IDs and client secrets in our characteristic study. 
The 98 credentials shown in Table~\ref{tab:RQ1Category} include client IDs and client secrets in our sampled dataset.
Among them, the counterparts of 53 credentials are also leaked in their host apps.
For example, Listing~\ref{listing:twitter_keys} shows a pair of Twitter client ID and client secret extracted during our characteristic study.
The secret (\texttt{TWITTER\_SECRET}) was included in our sampled dataset for manual inspection, and the corresponding ID (\texttt{TWITTER\_ID}) was stored just one line preceding the secret. 


Figure~\ref{fig:token_workflow} shows the workflow of the client credentials grant type of OAuth~\cite{oauth_client_credentials_type}.
In this case, the client ID and secret are used to issue API tokens that are later used to invoke other APIs provided by the service.
This process does not require user intervention and is typically used to access client-only resources.
Twitter provides application-only cloud APIs for third-party apps to access Twitter resources.
The Twitter credentials in Listing~\ref{listing:twitter_keys} can also be used with this workflow. 
 


\begin{figure*}
	\centering
	\includegraphics[width=\linewidth]{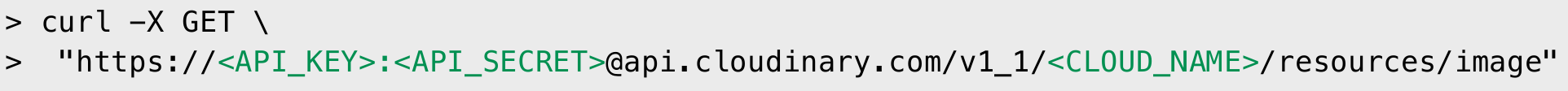}
	\captionof{lstlisting}{The Command to Get an App's Image Resources on Cloudinary Where \textit{\textless API\_KEY\textgreater}, \textit{\textless API\_SECRET\textgreater}, and \textit{\textless CLOUD\_NAME\textgreater} Should Be Changed Accordingly~\cite{cloudinary}.}
	\label{listing:cloudinary_command}
\end{figure*}

\begin{figure}
	\centering
	\includegraphics[width=0.9\linewidth]{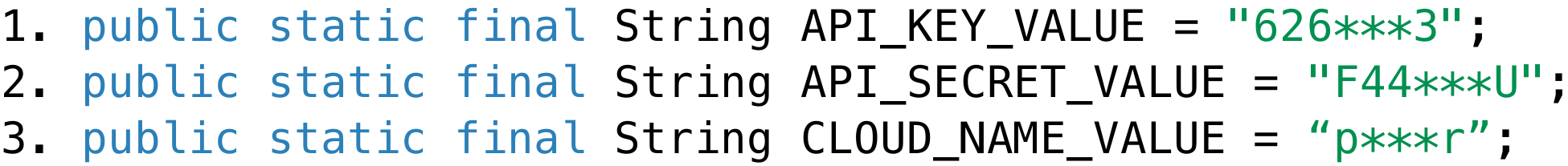}
	\captionof{lstlisting}{Anonymized Cloudinary Credentials Extracted in Our Dataset. The secrets were extracted from an app with over 1 million downloads on Google Play.}
	\label{listing:cloudinary_keys}
\end{figure}

We also found cloud service APIs that adopt Basic Authentication, i.e., directly using the app credentials for authentication.
For example, Cloudinary is a cloud service to host and optimize app media such as images and videos~\cite{cloudinary}.
The Admin APIs adopt basic authentication with commands such as the one shown in Listing~\ref{listing:cloudinary_command}.
It requires an \texttt{API\_KEY} and an \texttt{API\_SECRET} to authenticate the API and access the images of the app identified by the \texttt{CLOUD\_NAME}.
In our characteristic study, we found a pair of \texttt{API\_KEY} and \texttt{API\_SECRET} together with their corresponding \texttt{CLOUD\_NAME} as shown in Listing~\ref{listing:cloudinary_keys}.
The \texttt{API\_KEY} was included in our sampled dataset and its corresponding \texttt{API\_SECRET} and \texttt{CLOUD\_NAME} were defined right succeeding the \texttt{API\_KEY}.
With these leaked credentials, one can access, update, and even delete the app's media resources.
These credentials are meant to be used by back-end services and should be kept confidential. 

\footnotetext{For ease of understanding, we show the Java code in the examples of this paper. In our experiment, we worked on smali code.}

\textit{Public app keys.}
\begin{figure}
	\centering
	\includegraphics[width=0.9\linewidth]{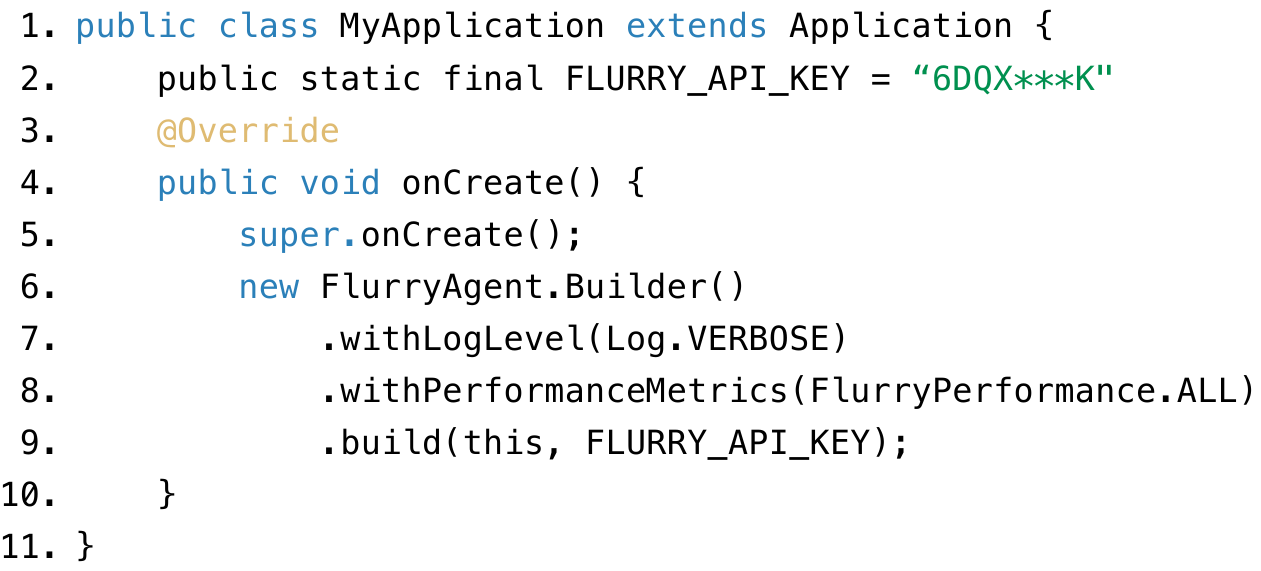}
	\captionof{lstlisting}{Anonymized Flurry API Key and Its Usage. The code snippet was adapted from an app with over 50 million downloads on Google Play.}
	\label{listing:flurry_keys}
\end{figure}
\begin{figure}[t]
	\centering
	\includegraphics[width=0.9\linewidth]{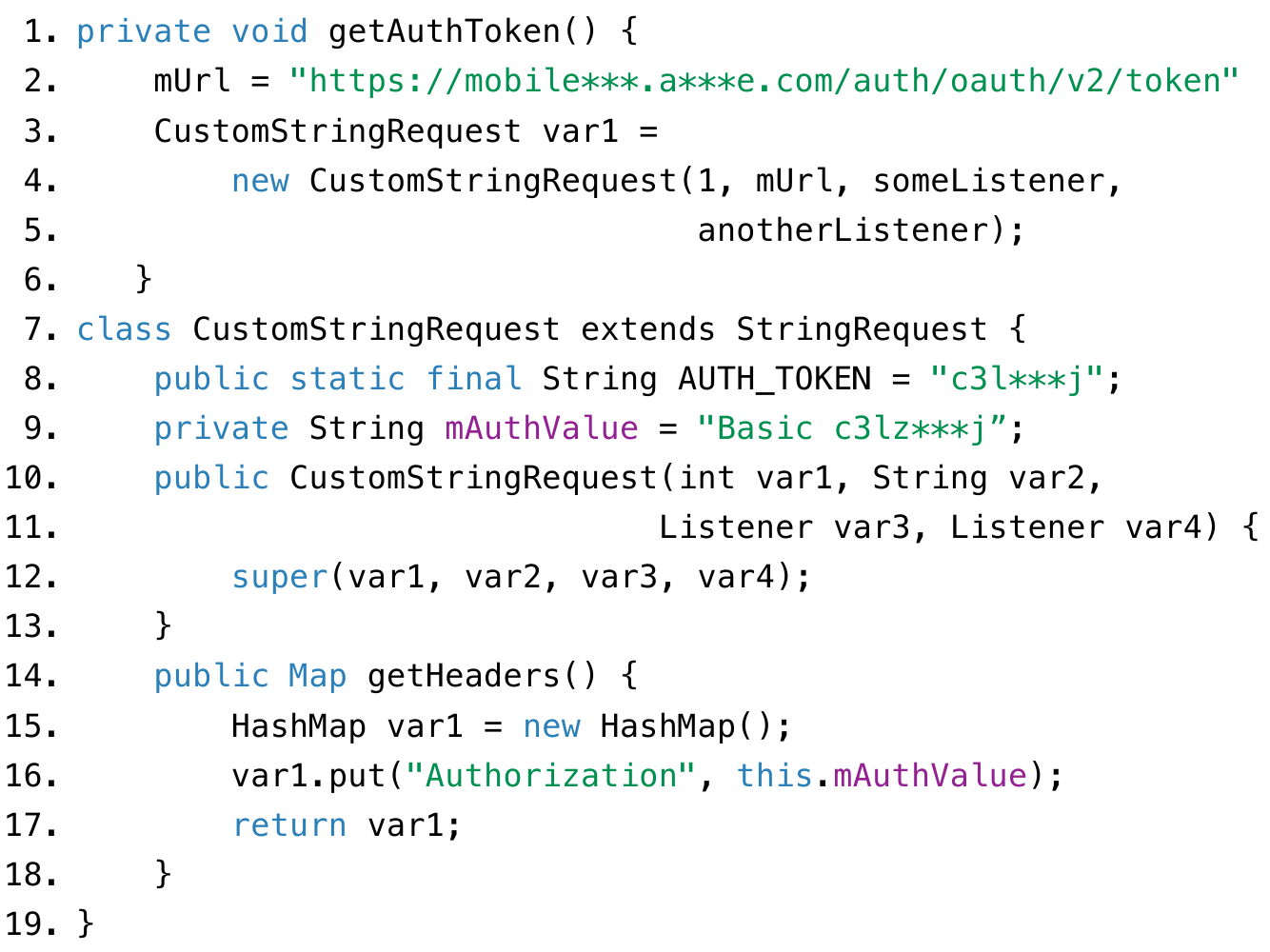}
	\captionof{lstlisting}{Adapted Code to Construct a Request for an App-private Back-end Service. The code was extracted from app com.a***e.a***b where the domain name in \texttt{mUrl} is the same as that in the app package name. The app has 50 thousand downloads on Google Play.}
	\label{listing:app_apecific_code}
\end{figure}
Apart from the sensitive app credentials, there are also 127 public app keys for different cloud services in our dataset. 
We define public app keys as those API keys that have limited permissions and are specified to be used in front-end clients in the API documents.
Different from the previously-discussed client IDs and \texttt{API\_KEY}s in multi-factor credentials, public app keys in this category cannot be paired with client secrets or \texttt{API\_SECRET}s for cloud service authentication.
We do not categorize these public app keys as irrelevant secret candidates because they also represent the identity of their host apps.
Typical examples in this category include app keys for push notification services, app keys for mobile analytics services, and app identifiers for advertising services.
For example, Flurry is a framework for mobile analytics and push notification services~\cite{flurry}.
To initialize Flurry SDK, an app needs to register with a valid API key as the example shown in Listing~\ref{listing:flurry_keys}.
This API key is meant to be used in the app as specified in the documentation~\cite{flurry}.
The other public keys are app identifiers for advertising services.
These identifiers are used to collect the ad traffic from an app.
These statistics are used to calculate the amount that an app should earn from the advertisers.


\textbf{App credentials for app-private back-end services.}
Apart from app credentials for third-party cloud services, we also identified 24 app credentials that are used in apps for authentication with app-private back-end services.
We categorized a secret candidate into this category if we found that it is used in authentication information for requests to URLs whose domain name is the same as the package name of the app, and we failed to identify any public documents indicating that the URLs are public cloud APIs.
Listing~\ref{listing:app_apecific_code} shows an example code snippet that constructs an HTTP request to an app-private back-end cloud service.
The requested URL shares the same domain name with the app package name (a***e.com) and our extracted secret candidate (\texttt{c3lz***j}) is used in the authorization header of the request (lines 8--9 and line 16).

\begin{figure}[t]
	\centering
	\includegraphics[width=\linewidth]{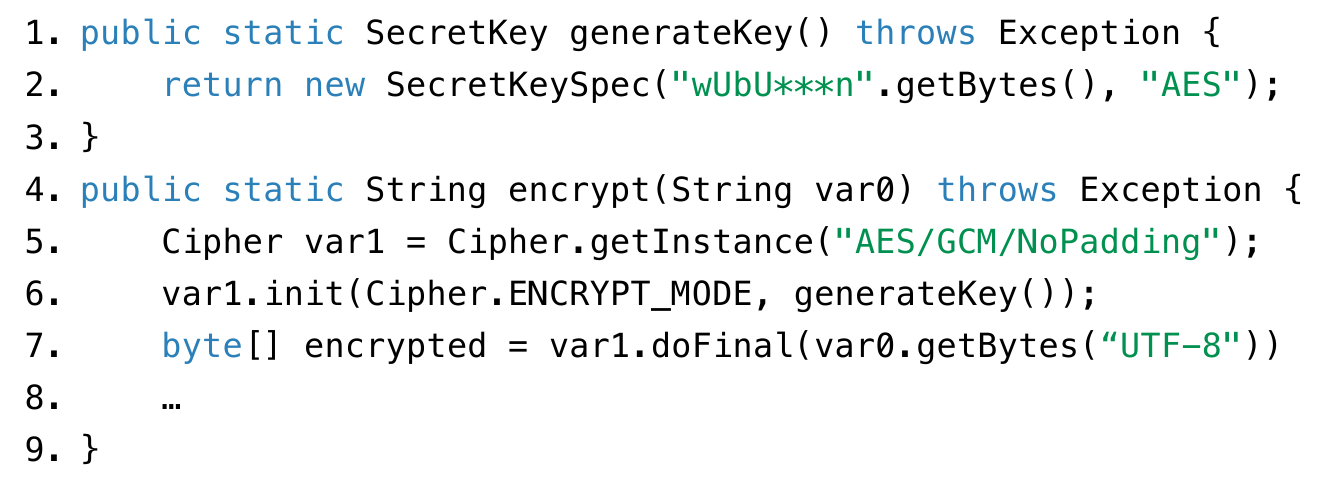}
	\captionof{lstlisting}{Adapted Code of an Encryption Helper Class. This class is used to encrypt and decrypt user sensitive information such as user names and passwords. The secret candidate in this example is the encryption key used in line 2, \texttt{wUbU***n}. We only show the encrypt method. There is also a decrypt method using the same key. The code snippet is adapted from an app with over 10 million downloads on Google Play.}
	\label{listing:encryption_example}
\end{figure}

\textbf{Encryption keys.}
Another 25 secrets are encryption keys.
They are used to encrypt or decrypt various data in Android apps.
We categorize a secret candidate into this category if we found that it was passed as keys to encryption APIs.
Using static encryption keys has been identified as a problem in Android for long~\cite{egele2013empirical}.
However, our study shows that this problem still occurs in popular Android apps and the static encryption keys can be easily extracted.
Listing~\ref{listing:encryption_example} shows a code snippet where the secret candidate (\texttt{wUbU***n} in line 2) is used to create a \texttt{SecretKeySpec} for AES algorithm.
It is then used to initiate a \texttt{Cipher} and encrypt the input string (lines 6--7).
In the same class, the same key is also used to decrypt strings.
For ease of presentation, we did not include the decrypt method in the listing.
This class is used to encrypt and decrypt user sensitive information such as user names and passwords that are later stored in \texttt{SharedPreferences}.

\textbf{Other secret candidates.} For the remaining secret candidates, 188 are obviously not app secrets such as keys for \texttt{SharedPreferences}~\cite{sharedPreferences}, a data structure in Android to store key-value pairs.
Twenty-one are app credentials for cloud services that are already not maintained.
As for the remaining 60 secret candidates, we cannot decide whether they are app secrets.
For example, there are a considerable number of secret candidates with sensitive variable names, but we failed to locate where the variables are used.
As a result, we cannot draw a conclusion whether these candidates are app secrets.
Since these cases do not share common patterns, we do not make further analysis on them.

\noindent\fbox{%
    \parbox{0.98\columnwidth}{%
        \textbf{Answer to RQ1:} \textit{Different types of app secrets can be leaked in Android apps including app credentials for third-party cloud services, app credentials for app-private back-end services, and encryption keys.
        Among them, app credentials for third-party cloud services account for the majority.
        Our results also demonstrate that app secrets can be extracted with simple approaches based on regular expressions.
        }
    }%
}

\subsection{RQ2 Secret Exploitation}\label{ssec:RQ2}
In this research question, we aim to study how the identified app secrets can be exploited and to understand the potential consequences of app secret leakage.
By answering this question, we assess the security impact of app secret leakage issues.
Different types of app secrets can have different security impact according to their usages.
As a result, we answer this research question based on our understanding of app secret usages learned in RQ1.
Specifically, we tracked how the secrets were used (e.g., by which API calls) and derived the exploitation approaches accordingly.
Below, we discuss the exploitation approaches for different types of app secrets.


\textbf{Impersonating the victim apps and invoke third-party cloud APIs.}
As discussed in RQ1, many of the leaked app secrets are third-party cloud service credentials used to authenticate cloud service API invocations (e.g., Listing~\ref{listing:google_maps_command}, Listing~\ref{listing:cloudinary_command}, and Figure~\ref{fig:token_workflow} all describe cases where credentials are used to authenticate cloud API invocations).
When these credentials are leaked, attackers can make calls to the corresponding APIs while authenticated as the victim apps.
For single-factor credentials, an attacker can launch attacks with merely one credential (e.g., the \texttt{GOOGLE\_API\_KEY} in Listing~\ref{listing:regex_example} can be used to invoke Geolocation API with the command in Listing~\ref{listing:google_maps_command}).
As for multi-factor credentials, an attacker can only launch attacks when all the required credentials are leaked (e.g., the values of \texttt{API\_KEY}, \texttt{API\_SECRET} and \texttt{CLOUD\_NAME} in Listing~\ref{listing:cloudinary_keys} are all needed to invoke Cloudinary APIs as shown in Listing~\ref{listing:cloudinary_command}).

Leaking such credentials can induce various consequences such as monetary losses, denial of service (DoS), and leakage or damage of app-private information.
Many of the cloud APIs are priced according to the number of calls to them.
In our study, we found that calls to APIs of Google Maps, Google Cloud Vision, Google Translation, Azure Speech Service API, etc. all follow such a pricing plan.
For example, the price of Static Street View API in Google Maps is priced at \$7 per 1000 requests.
An attacker can use the API with the victim app's leaked key while the app needs to pay the bill.
In addition, invocations to cloud APIs commonly adopt a rate limit.
For example, many of the standard APIs of Twitter have rate limits with regard to each app~\cite{twitter_rate_limit}.
The API to show the recent tweets liked by a user is bounded at a rate of 75 requests per 15 minutes~\cite{twitter_rate_limit_favorites}.
An attacker can launch DoS attacks by continuously sending requests with the credentials of a victim app to invoke the API to exceed its rate limit.
Consequently, normal API invocations made by the victim app and its users will be blocked.
Leaking cloud API credentials can also induce leakage or damage of app-private data.
For example, the Cloudinary API key and secret shown in Listing~\ref{listing:cloudinary_keys} can be used to access, update and delete the app's media resources hosted on Cloudinary.
An attacker can manipulate the resources once they have obtained the key and the secret.

\textbf{Impersonating the victim apps and phishing for user access tokens.}
As shown in Figures~\ref{fig:oauth} and~\ref{fig:token_workflow}, app credentials can be used for OAuth with different workflows.
For those cloud APIs authenticated with client credentials grant type, an attacker can directly use the leaked credentials to invoke the corresponding APIs.
OAuth credentials can be further exploited by making use of the authorization code grant type.
As the authorization code grant type involves users granting access to the apps, an attacker can launch phishing attacks by impersonating the victim apps.
Specifically, the attacker can build a fake app with the victim app's credentials to replace the ``Client'' in Figure~\ref{fig:oauth} to communicate with the cloud service and the user.
In such a phishing attack scenario, both the user and the cloud service will identify the fake app as the victim app as it holds the victim apps' credentials.
Once the attacker obtained the access token, he can access or manipulate the victim users' private information.


\textbf{Exploiting encryption keys.}
We found 25 encryption keys that are embedded in the apps and stored as plain text.
They are used to encrypt and decrypt different types of data.
The encryption key in Listing~\ref{listing:encryption_example} is used to encrypt user sensitive information that are saved in \texttt{SharedPreferences}.
We also found leaked encryption keys used to decrypt the information obtained when scanning QR codes.
Leakage of encryption keys allows attackers to decrypt the encrypted information, making the encryption meaningless~\cite{egele2013empirical}.

\textbf{Exploiting app credentials for app-private back-end services.}
In our characteristic study, we also identified 24 app credentials for app-private back-end services which can be exploited to access the private services of the victim apps.
However, since different apps have different protocols to access their back-end services, it is not easy to derive common exploitation approaches for these app-private credentials.
In this paper, we focus on general app secrets that can occur in different apps, and leave exploiting app-private secrets as our future work.

\textbf{Exploiting public app keys for analytics services.}
Many of our extracted public app keys are app credentials for analytics services, which are meant to be used in Android apps.
These credentials are used when submitting user behavioral data to the cloud service to identify the concerned apps.
Although they are specified as public app keys by the cloud service providers, attackers may also abuse them to spam the victim apps by sending fake user behavior.
Whether this kind of attack is exploitable depends on the intrusion prevention techniques deployed by the analytics services.
Powerful intrusion prevention techniques may have been deployed to identify and remove the fake data.
However, since the details of such intrusion prevention techniques deployed by various analytics services are mostly unavailable, it is difficult for us to assess whether a key can be truly exploited.
Intrusion and intrusion prevention techniques can be another research direction~\cite{fuchsberger2005intrusion}.

\noindent\fbox{%
    \parbox{0.97\linewidth}{%
        \textbf{Answer to RQ2:} \textit{
        In general, the leak of app credentials of cloud services allows attackers to impersonate victim apps and steal their privilege, and the leak of encryption keys allows attackers to decrypt the protected information.
        Consequently, leaking app secrets can induce various security issues such as monetary losses, denial of service, and leakage or damage of private information.
        }
    }%
}

\subsection{RQ3 Bad Practices}
While checking the data entries in our characteristic study, we observed several bad practices that make it easy for attackers to steal app secrets.
In this section, we discuss the bad practices we observed.

\textit{\textbf{Observation 1: } App secrets are hard-coded in Android apps as plain text and their variable names are not obfuscated.}

As discussed in Section~\ref{ssec:data_collection}, we leveraged the regular expression in Listing~\ref{listing:regex} to search for app secrets.
Essentially, the regular expression searches for hard-coded app secrets whose names contain sensitive keywords.
Our successful identification of app secrets with this procedure means that there are hard-coded app secrets whose variable names are not properly obfuscated.
This is a bad practice since an attacker can easily reverse-engineer Android apps and extract the secrets.
This also suggests the potential to leverage such contextual information to develop machine learning techniques to detect hard-coded secrets.
The use of hard-coded credentials has been well recognized as a security issue~\cite{cwe321,cwe798}.
App developers should avoid using hard-coded app secrets in their apps.

\textit{\textbf{Observation 2: } Apps tend to store different secrets in the same file.
227 of our studied app secrets were extracted from files that contain multiple secrets.}

We inspected the files from which the secret candidates were extracted.
Our inspection found that many of these files contained app secrets in addition to the ones that were inspected in our characteristic study.
227 of our studied app secrets were extracted from files that contain multiple secrets.
For example, the app credentials shown in Listing~\ref{listing:twitter_keys} are stored in a class named \texttt{Keys}.
Besides these Twitter app credentials, there are also app credentials for WeChat, Flurry and other cloud services.
It is unsafe to put all kinds of app secrets in one file.
It allows an attacker to easily obtain many app secrets, even for multi-factor app credentials, from a single file.
Automatic tools can be designed to extract multi-factor credentials by enumerating the combination of the strings within the same files.

\textit{\textbf{Observation 3: } App secrets that are not needed by client apps as specified by the cloud service documents can also be included in apps.}

\begin{figure}[t]
	\centering
	\includegraphics[width=\linewidth]{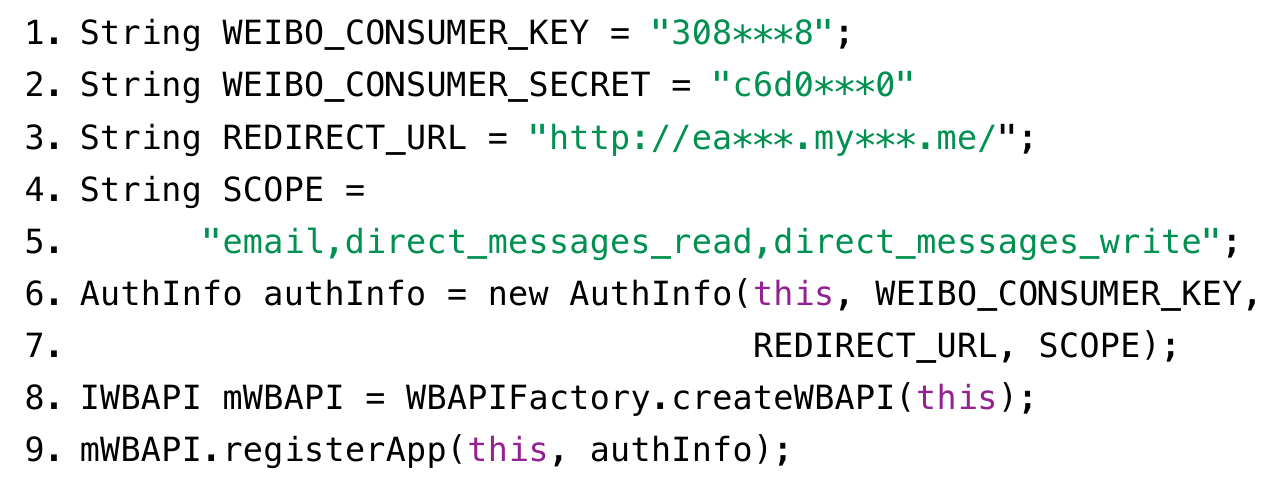}
	\captionof{lstlisting}{OAuth with Weibo Using Weibo SDK. While the SDK does not require the client secret (lines 6--9), the app still embed the secret (line 2). The code snippet is adapted from an app with over 50 thousand downloads on Google Play.}
	\label{listing:weibo_example}
\end{figure}

To protect app credentials, some cloud services like Weibo and WeChat implement their Android SDK in such a way that client secrets are not needed in client apps.
However, we still found 20 Weibo and WeChat client secrets embedded in apps.
For example, Listing~\ref{listing:weibo_example} shows a code snippet to initialize Weibo SDK for OAuth with Weibo.
The OAuth process implemented by the SDK does not require the client secret (line 6--9) but the app still embedded the secret (\texttt{WEIBO\_CONSUMER\_SECRET} in line 2).
This leads to unnecessary leakage of the client secret.


\textit{\textbf{Observation 4: } Cloud services like Google Maps support applying restrictions to app credentials. However, many apps did not restrict their credentials.}

We observed that many of the cloud services allow developers to set restrictions on their app credentials.
For example, Google Cloud Services, such as Google Maps and Google Translation, allow developers to specify the app that can be used with the API keys by restricting the app signature.
However, setting restrictions is not mandatory and as a result, some apps did not restrict their app secrets.
We found four valid and usable Google Cloud Service keys, i.e., these keys are not restricted.
In Section~\ref{sec:detection}, we show that thousands of unrestricted Google Cloud Service keys can be extracted from popular apps on Google Play.

\noindent\fbox{%
    \parbox{0.97\linewidth}{%
        \textbf{Answer to RQ3:} \textit{
        Apps may not properly protect their app secrets.
        In our characteristic study, we observed several bad practices in storing app secrets. These secrets are hard-coded and stored in a single file without obfuscation.
        There are also unused app secrets that are still included in the apps.
        Furthermore, app developers may not restrict their app secrets.
        As a result, an attacker can easily steal and exploit app secrets in Android apps.
        App developers should avoid such bad practices and better protect their secrets.
        }
    }%
}


\subsection{Implications}
\textbf{App secrets are not far from being stolen.}
Many app secrets can be stolen using a simple keyword search.
In our characteristic study, we have shown that hundreds of app secrets can be extracted from Android apps by regular expression matching and manual inspection.
Since we just inspected a small portion of the regular expression matching results, there can be a lot more app secrets beyond our inspection.
Using hard-coded secrets is a well-recognized security issue~\cite{cwe321,cwe798}.
It is surprising that a large number of Android apps still have this problem.
All the apps used in our characteristic study are the most popular ones under each Google Play category.
Despite their popularity, many apps failed to properly protect their secrets.
This suggests a general lack of awareness of the problem. It calls for stronger measures to be taken by developers to protect their app secrets.


\textbf{Protecting app credentials of cloud services.}
It is difficult to guarantee the safety of cloud service secrets at the app end. Nevertheless, measures can be taken to make the secrets harder to extract.
As shown in our characteristic study, we can extract a considerable amount of app secrets by keyword matching on variable names.
Obfuscation helps defend such simple attacks by obscuring the variable names.
To prevent more sophisticated attacks, an app can encrypt the secrets or maintain back-end services to dispatch the secrets.
With such measures, it is difficult to extract cloud service secrets statically.
However, if an attacker crafts a delicate dynamic analysis, cloud secrets can still be exposed.

To better protect the secrets, mechanisms at the cloud end can be deployed.
For example, some cloud services allow app developers to set restrictions on the key.
App developers can configure their API keys only used with APK files of certain signatures.
We observed such mechanisms supported by popular cloud services, including Google Cloud services, Weibo, and WeChat.
However, the restrictions are not necessarily compulsory, such as those in Google Cloud services.
Although the API document suggests the developers to restrict their keys~\cite{googlemapsapirestriction}, there are still many apps not following the suggestion.
The cloud services may better alert the app developers when they did not restrict their secrets.

\textbf{Protecting encryption keys.}
In the cases where hard-coded static encryption keys are needed, the approaches discussed above to protecting cloud service secrets apply.
For example, obfuscation and secret dispatch from server are also useful to hide the encryption keys.
In some scenarios, encryption keys can be non-static.
The encryption key in Listing~\ref{listing:encryption_example} is an example of such a scenario.
The keys are used to encrypt and decrypt information that is only stored locally.
In this case, encryption keys can be randomly generated.
Android provides a \texttt{KeyStore} system to randomly generate keys and securely store them.
Every time an app needs to encrypt or decrypt information, the app can retrieve the stored encryption keys from \texttt{KeyStore}.
It provides a more secure key management than hard coding the encryption keys.

    \textbf{Google Play policies can boost the secure storage of app secrets.}
    Google Play scans APK files and reports potential issues, including hard-coded secrets, before apps are published.
    It provides checks for specific types of app secrets, and our results show that despite this scanning, apps with hard-coded secrets can still be published.
    Notably, Google Play has enforced a policy that prevents apps with hard-coded AWS credentials from being published~\cite{googleplayaws, stackoverflowgoogleplay}.
    Once detected with hard-coded AWS credentials, apps will be removed from Google Play if the exposed credential is not removed from the app before a specified deadline.
    In our manual analysis, we identified a leaked AWS credential in a weather app from 2017, but subsequent versions of the app no longer contained this credential.
    This case demonstrates that Google Play policies can enhance the secure storage of secrets in apps hosted on their platform.
    To mitigate the problem of hard-coded secrets, app stores like Google Play may play a stronger role by improving the effectiveness of hard-coded secret detection or enforcing stricter policies.
    Nonetheless, the design of the policies also require delicate considerations.

\section{Detecting Exploitable Cloud Service Credentials}\label{sec:detection}
\begin{table*}[]
\caption{\centering Regular Expressions to Detect Exploitable Third-Party Cloud Service Credentials.}
\label{tab:credential_detection}
\centering
\resizebox{\textwidth}{!}{
\begin{tabular}{c|c|c}
\hline
\textbf{Cloud Service}                                                                                   & \textbf{Credential} & \textbf{Regular Expression}                                                              \\ \hline
\begin{tabular}[c]{@{}c@{}}Google Maps\\ Google Translation\\ Google Cloud Vision\\ YouTube\end{tabular} & API key             & AIza{[}0-9A-Za-z\_-{]}\{35\}                                                             \\ \hline
FCM                                                                                                      & API key             & AAAA{[}A-Za-z0-9\_-{]}\{7\}:{[}A-Za-z0-9\_-{]}\{140,162\}|AIzaSy{[}0-9A-Za-z\_-{]}\{33\} \\ \hline
\multirow{2}{*}{Facebook}                                                                                & Client ID           & {[}0-9{]}\{13,17\}                                                                       \\ \cline{2-3} 
                                                                                                         & Client secret       & {[}0-9a-f{]}\{32\}                                                                       \\ \hline
\multirow{2}{*}{Twitter}                                                                                 & Client ID           & {[}0-9a-zA-Z{]}\{18,25\}                                                                 \\ \cline{2-3} 
                                                                                                         & Client secret       & {[}0-9a-zA-Z{]}\{40,50\}                                                                 \\ \hline
\end{tabular}
}
\end{table*}
In RQ2, we make observations of the exploitation approaches for different kinds of app secrets and discuss their potential impacts.
However, it is still unclear whether the leaked secrets are truly exploitable and how prevalent the exploitable app secrets are.
To investigate this question and understand the impact of app secret leakage in the wild, we conducted a study to automatically detect exploitable app credentials of several cloud services including Google Maps, Google Translation, Google Cloud Vision, YouTube, Firebase Cloud Messaging, Twitter, and Facebook.
We are interested in exploiting third-party cloud service credentials because they are the most prevalent kind of app secrets in our characteristic study (Table~\ref{tab:RQ1Category}).
We choose these cloud services because (1) they are the most popular, (2) their credential leaks are likely to have security impacts, 
and (3) their credentials can be validated by making cloud API calls without human intervention.

This study complements our manual analysis in Section 4 by providing a broader picture of the prevalence of exploitable leaked credentials in Android apps through the automatic detection of certain types of app secrets. It offers quantitative evidence for the prevalence of leaked secrets in Android apps.
\subsection{Approach}
To detect the cloud service credentials, we borrow the idea adopted by Meli et al.~\cite{meli2019bad}.
We define a regular expression for each type of the cloud service credentials based on their instances observed in our characteristic study.
The regular expressions are summarized in Table~\ref{tab:credential_detection}.
Different from the approach adopted by Meli et al.~\cite{meli2019bad}, we also validate the extracted credentials by using the credentials to invoke APIs in the corresponding cloud services and checking the response.
We describe our approach to detect and validate the credentials of each cloud service below:

\textbf{Google Maps, Google Translation, Google Cloud Vision, and YouTube.}
We introduce these services collectively because they are all owned by Alphabet Inc. and share the same format.
The APIs of these four services require single-factor API keys for authentication such as the one shown in Listing~\ref{listing:google_maps_command}.
The API keys of these services are in the same format: they (1) start with ``AIza'', (2) are 39 characters long, and (3) only contain letters, numbers and underscores (Table~\ref{tab:credential_detection}).
Leaking their API keys can induce monetary losses or DoS.
Google Maps, Google Translation and Google Cloud Vision APIs all adopts pay-as-you-go pricing plans.
The costs are calculated based on API usage.
App developers can also set rate limits to the API keys.
In such scenarios, attackers can also launch DoS attacks to exceed the API rate limits.
As for YouTube, it is free to invoke YouTube APIs with rate limits.
Attackers can launch DoS attacks with leaked YouTube API keys.

\textbf{Firebase Cloud Messaging (FCM).}
Firebase Cloud Messaging (FCM) is a service to send push notifications to Android apps.
Leaking FCM server keys will allow attackers to send arbitrary notifications to victim apps.
We adapted the regular expressions published by Abss~\cite{FCM_takeover} to detect FCM server keys.
Specifically, Abss used a regular expression to extract fixed-length strings (\texttt{AAAA[a-zA-Z0-9\_-]{7}:[a-zA-Z0-9\_-]\{140\}}).
In our characteristic study, we observed that the length of FCM server keys can vary so we adapted the regular expression to search for strings whose length is within the range (as shown in Table~\ref{tab:credential_detection}).
Note that Abss observed deprecated Google Cloud Messaging (GCM) keys can also be used for FCM notification services and GCM keys share the same pattern with Google API keys.
As a result, we also leveraged the regular expression to search for Google API keys to search for FCM server keys.

\textbf{Twitter and Facebook.}
Different from previous cloud services, app credentials for Twitter and Facebook are multi-factor.
Both services require a client ID and a client secret for service authentication.
In addition, their credentials do not exhibit strong patterns that allow one to distinguish them.
For example, Facebook client IDs are sequences of numbers with lengths ranging from 13 to 17 (Table~\ref{tab:credential_detection}).
Using merely regular expressions to extract them will result in too many noises.
The noises can significantly slow down the validation process since client IDs and client secrets need to be combined for validation.
$10^3$ noisy client IDs and $10^3$ noisy client secrets will result in $10^6$ times of validation.
As a result, it is crucial to reduce noises for these two services.
We combined both keyword matching and regular expression matching to reduce the noises in extracting app credentials for Twitter and Facebook services.
Specifically, we extract a string value for further validation if (1) its value matches the regular expression, (2) its variable name contains keywords in Listing~\ref{listing:regex}, and (3) its variable names or filenames contain the corresponding cloud service name or its acronyms (e.g., Facebook and FB).

Whenever we identified a potential client ID or client secret using these filters in a file $f$, we searched for its counterparts only in $f$.
For example, if we have extracted a string that is a potential Facebook client ID by matching the regular expression and the keywords, we locate the file and use only the regular expression to extract potential Facebook secrets.
In this way, we can find more pairs of Twitter and Facebook credentials in case one of them does not follow the naming convention.
This additional step helped extract Facebook credentials ten times more.

Exposing Twitter or Facebook credentials may lead to different security consequences, as both platforms offer a variety of services and different credentials can have varying access permissions. To avoid undesirable ethical concerns, we only validated whether the detected Twitter and Facebook credentials could successfully invoke the authentication APIs. Further exploring the specific permissions of each detected credential could result in undesirable consequences. For example, evaluating whether a pair of Facebook credentials can submit posts would require invoking the corresponding APIs and making actual submissions, which clearly violates our ethical considerations.

\textbf{Validating Extracted Credentials.}
Not all the strings matching the regular expressions are valid credentials. There can exist irrelevant strings that coincidentally match with the regular expressions or example credentials that are not exploitable.
To address this problem, we validate the extracted credentials by using it to invoke APIs of each cloud service and checking the response code.
We consider the credentials as valid ones if cloud service API invocations using them return 200 (OK Success)~\cite{httpreturncode}.
We built a validator for each of the cloud services.
Checking the return code can determine whether a string is a valid credential since all of our detected credentials are used for authorization. The APIs require valid credentials to be accessed, and they will return error codes such as 403 Forbidden if the credentials are not valid.
This ensures that all the validated credentials are exploitable.
For Google Maps, we adapted the validator published by Ozgur Alp~\cite{google_maps_key_blog}.
For YouTube, we adapted the validator provided by KeyHacks~\cite{keyhacks}.
For Google Cloud Vision and Google Translation we implemented our own validators.
For Twitter and Facebook, we used their APIs to issue app tokens for validation.
We combined each of the extracted client IDs with the client secrets and used the pairs to invoke the token issuing APIs.
We only validate client IDs and client secrets extracted from the same app to mitigate the combination explosion problem.
It is unlikely that a pair of valid multi-factor credentials for one app are included in different apps.
We consider a pair of credentials as valid Twitter or Facebook credentials if we can successfully generate tokens with the pair. We did not further exploit the tokens for ethical consideration.
The validators are available on our project website~\cite{homepage}.


\textbf{Dataset.}
While our manual analysis and conclusions in Section 4 are drawn from a specific dataset, solely using our proposed regular expressions on the same dataset raises concerns of overfitting.
To mitigate the problem, we introduced a new dataset in this study.
We collected the top 100 apps that are most popularly downloaded in each category of Google Play in 2021.
We successfully downloaded 8,376 apps.
Some apps could not be downloaded because of district or device restrictions.
We detected exploitable third-party cloud service credentials in both the app dataset used in our characteristic study (Section~\ref{ssec:data_collection}) and in the newly collected dataset.
In total, we analyzed 23,041 apps.
With the apps collected from different years, we were also able to conduct a temporal analysis to analyze the life spans of the detected secrets.

\textbf{Discussions.}
Our approach to detecting exploitable third-party credentials can have false negatives due to many reasons.
For example, single-factor credentials that are calculated by string computations could be missed due to the inappropriateness of the regular expressions used.
Twitter and Facebook credentials with obfuscated variable names could also be missed.
In addition, we based our assessment on only several most popular cloud services.
There are far more additional cloud services that are used by Android apps whose credentials can be leaked and exploited.
As a result, there can be considerably more exploitable app secrets leaked in the Android apps.

\subsection{Results}
\begin{table}[]
\centering
\caption{Number of Exploitable Credentials. For multi-factor credentials, the value shows the number of exploitable credential pairs. Column Total shows the total number of distinct exploitable credentials in both datasets. Overlap shows the number of credentials that can be extracted from both datasets.}
\label{tab:detection_results}
\begin{tabular}{@{}lrrrr@{}}
\toprule
\textbf{Cloud Service} & \textbf{2017--2020} & \textbf{2021} & \textbf{Total} & \textbf{Overlap} \\ \midrule
Google Maps            & 1,886               & 1,094         & 2,080          & 900              \\
Google Translation     & 354                 & 329           & 475            & 208              \\
Google Cloud Vision    & 287                 & 265           & 366            & 186              \\
YouTube                & 227                 & 183           & 272            & 138              \\
Twitter                & 208                 & 147           & 256            & 99               \\
FCM                    & 197                 & 98            & 223            & 72               \\
Facebook               & 23                  & 30            & 39             & 14               \\\midrule
Total                  & 3,182               & 2,146         & 3,711          & 1,617            \\\bottomrule
\end{tabular}%
\end{table}

\begin{table*}[]
\caption{Top 10 Apps with the Most Number of Exploitable Credentials. In our dataset, there are 15 apps with five exploitable credentials. We show the most popular three in this table. Some old apps are no longer available on Google Play. We mark their ratings and downloads with ``-''.}
\label{tab:apps_most_credentials}
\centering
\resizebox{\textwidth}{!}{%

\begin{tabular}{@{}lllllr@{}}
\toprule
\textbf{App ID}           & \textbf{Year} & \textbf{Rating} & \textbf{\#Downloads} & \textbf{Services}                                                                                      & \textbf{\#Credentials} \\ \midrule
App-1: c***.m***.t***     & 2017             & -               & -                 & \begin{tabular}[c]{@{}l@{}}Google Maps\\ FCM\end{tabular}                                             & 9                      \\ \midrule
App-2: c***.g***.a***.*** & 2021             & 4.1             & 5,000,000,000+    & \begin{tabular}[c]{@{}l@{}}Google MapsGoogle Translation\\ Google Cloud Vision\\ YouTube\end{tabular} & 7                      \\ \midrule
App-3: b***.c***.s***.*** & 2021             & 4.6             & 1,000,000+        & Google Maps                                                                                           & 7                      \\ \midrule
App-4: c***.v***.a***     & 2017             & 3.9             & 10,000,000+       & \begin{tabular}[c]{@{}l@{}}Google Maps\\ Google Translation\\ Google Cloud Vision\\ FCM\end{tabular}  & 6                      \\ \midrule
App-5: c***.G***.G***     & 2018             & -               & -                 & \begin{tabular}[c]{@{}l@{}}Google Maps\\ Google Translation\end{tabular}                              & 6                      \\ \midrule
App-6: c***.m***.g***.*** & 2018             & -               & -                 & \begin{tabular}[c]{@{}l@{}}Google Maps\\ Google Translation\end{tabular}                              & 6                      \\ \midrule
App-7: c***.t***.t***     & 2019             & 4.4             & 100,000,000+      & Google Translation                                                                                    & 6                      \\ \midrule
App-8: c***.p***.s***     & 2019             & 4.2             & 500,000,000+      & \begin{tabular}[c]{@{}l@{}}YouTube\\ Twitter\\ Facebook\end{tabular}                                  & 5                      \\ \midrule
App-9: c***.g***          & 2021             & 4.4             & 50,000,000+       & \begin{tabular}[c]{@{}l@{}}Google Maps\\ YouTube\\ FCM\end{tabular}                                   & 5                      \\ \midrule
App-10: b***.m***.c***    & 2020,2021        & 4.7             & 10,000,000+       & \begin{tabular}[c]{@{}l@{}}Google Maps\\ FCM\end{tabular}                                             & 5                      \\ \bottomrule
\end{tabular}%
}
\end{table*}

We ran the analysis in a total of 23,041 apps, among which 4,020 contains at least one exploitable app secret.
Table~\ref{tab:detection_results} summarizes the exploitable credential detection results.
For multi-factor credentials, we consider the multiple factors in a credential holistically. For example, each Twitter credential consists of a client ID and a client secrete. Therefore, 208 exploitable Twitter credentials refer to 208 pairs of exploitable Twitter client ID and client secret.
We split the results extracted in the dataset used in our characteristic study (i.e., dataset 2017--2020) and the newly collected dataset (i.e., dataset~2021).
We identified 3,182 exploitable cloud service credentials in dataset 2017--2020, and 2,146 in dataset 2021.
In total, we identified 3,711 distinct exploitable credentials where 1,617 of them (43.6\%) appeared in both datasets.

Among the apps that contain exploitable credentials, most of them contain only one instance (2,774 out of 4,020).
There are also apps containing multiple exploitable credentials.
Table~\ref{tab:apps_most_credentials} shows the top ten apps with the most exploitable credentials.
App-1 contains nine distinct exploitable credentials and is the app with the most exploitable credentials.
Another two apps contain seven exploitable credentials, and four apps contain six.
In our dataset, there are in total 15 apps containing five credentials.
In Table~\ref{tab:apps_most_credentials}, we only show three of them with the most number of downloads and the highest ratings.
As shown in the table, these apps are highly rated popular apps.
App-1, App-5 and App-6 are no longer available on Google Play. So we omitted their ratings and number of downloads.
All the other apps have at least 1 million downloads.
The most popular app (App-2) has over 5 billion downloads.
This shows that even popular apps can suffer from app secret leakage issues, and they may leak multiple app secrets at the same time.


\begin{table}[]
\centering
\caption{Exploitable App Credentials Contained in Over 10 Apps}
\label{tab:credentials_in_mutli_apps}
\begin{tabular}{@{}llr@{}}
\toprule
\textbf{Credential}            & \textbf{Cloud Service} & \textbf{\#Apps} \\ \midrule
Credential-1: AIza***          & Google Maps            & 53              \\
Credential-2: 2vvZ***, Ix1Y*** & Twitter                & 40              \\
Credential-3: AIza***          & Google Maps            & 39              \\
Credential-4: AIza***          & Google Maps            & 21              \\
Credential-5: AIza***          & YouTube                & 21              \\
Credential-6: AIza***          & Google Translation     & 18              \\
Credential-7: AIza***          & Google Maps            & 18              \\
Credential-8: AIza***          & FCM                    & 18              \\
Credential-9: AIza***          & Google Maps            & 17              \\
Credential-10: AIza***         & FCM                    & 11              \\
Credential-11: AIza***         & YouTube                & 11              \\ \bottomrule
\end{tabular}%
\end{table}

\textit{\textbf{Observation 5:} The same exploitable credentials can be used in multiple Android apps.}

1,775 of the 3,711 distinct credentials are only used in one app.
The other credentials are used in multiple apps.
Table~\ref{tab:credentials_in_mutli_apps} shows the exploitable credentials that are used in more than ten apps.
We manually inspected the apps that contain these credentials.
Some credentials are used in the same libraries that are used in different apps.
This applies to Credential-1, Credential-6, Credential-7 and Credential-8.
We made this observation based on the package names of the classes that contain the credentials.
Credential-1 is used in a library provided by an online app builder. The library is used by 53 different apps.
Credential-6, Credential-7 and Credential-8 are all used in a library of a weather provider and the apps containing these credentials are the same set of apps in Weather category.
As for Credential-9, all the 17 apps are included in the set of apps where Credential-6, Credential-7 and Credential-8 are extracted.
We suspect that Credential-9 is provided by the same library, but we cannot be sure since Credential-9 is defined in \texttt{AndroidManifest.xml} with common variable names.
This shows that popular libraries can also leak app secrets.
The libraries should properly protect the secrets since their leakage can affect a lot of apps.

Another four credentials were extracted from different apps developed by the same company (Credential-2, Credential-5, Credential-10 and Credential-11).
We recovered the company information by referring to the developer information on Google Play.
For example, all the 40 apps that contain Credential-2 are developed by the same company and the app package IDs share the same prefix.
This indicates that the same company can use the same set of app credentials in their different products.
However, this is a risky practice since once a credential is compromised and exploited by attackers, the impact can be large.
For example, if an attacker launches DoS attacks with the leaked credentials, all the apps of the company will be affected.
It is also more difficult to deal with the attacks when different apps share the same credentials.
When the app company wants to disable a compromised credential, it needs to take into account the users of all its apps.
It may need to urge all the users to upgrade their apps to a newer version before the credentials can be disabled.
This is definitely more difficult when more apps share the same comprised credential.

We failed to observe any patterns for the apps that contain Credential-3 and Credential-4.
Credential-3 is mostly used in apps in Dating and Beauty categories, and Credential-4 is used in calling and messaging apps.
However, the apps are developed by different developers, and we do not observe any relations among them.
This may suggest the existence of credential abuse.
\begin{figure}[t]
	\centering
	\includegraphics[width=\linewidth]{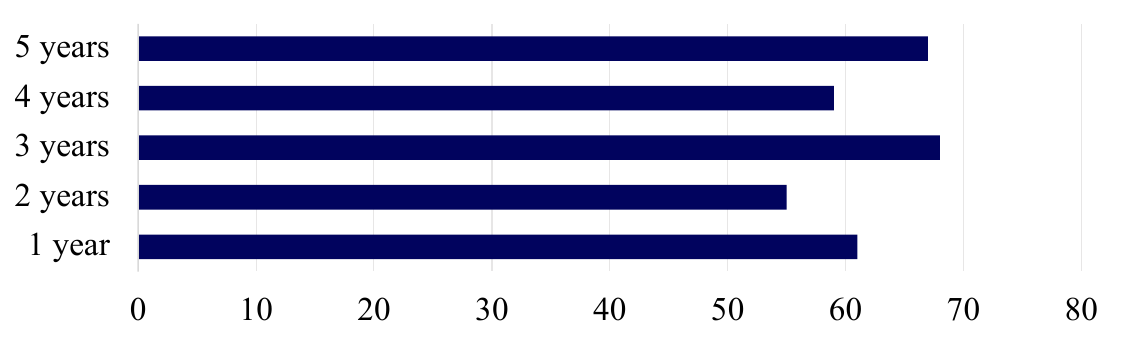}
	\caption{Life Spans of Exploitable Credentials in Apps That Are Contained in All Five Years.}
	\label{fig:credential_lifespan}
\end{figure}

\textit{\textbf{Observation 6:} The life span of exploitable credentials can be long. 21.6\% of our studied credentials were in their host apps for at least five years.
31.6\% of our studied credentials were removed from the latest version of the host apps while remaining exploitable.
}

In our collected app datasets, 378 apps are contained in datasets of all the five years (2017--2021).
We analyzed the exploitable app secrets in these 378 apps over the five-year span.
From these apps, we extracted 310 exploitable app secrets.
Figure~\ref{fig:credential_lifespan} shows the life spans of these credentials.
Among them, 67 (21.6\%) app secrets are extracted from all the five versions of the same app over the five years.
59 (19.0\%) have been present for four years, 68 (21.9\%) three years, 55 (17.7\%) two years and 61 (19.7\%) one year.
In other words, 80.3\% of the app credentials can be extracted from the same apps for at least two years.

We also observed that 98 (31.6\%) app credentials are present in an app's older versions, but are no longer included in 2021 version.
Among them, 20 are only available in the apps' 2017 versions.
Although these secrets are removed from the latest version, they are still exploitable.
The app developers need to disable these secrets since they are still subject to leakage if attackers obtain the old versions of their apps and many third-party websites provide old versions of apps for downloading~\cite{Aptoide,AndroidAPKsFree}.



\subsection{Developer's Perspectives on App Secret Leakage}
We reported our identified exploitable app secrets in dataset 2021 to their original app developers.
For a better chance of getting responses from developers, we reported only the detected app secret leaks in apps with over 100 million downloads and with public security issue reporting channels.
In total, there are 50 apps matching our selection criteria.
We divided the apps according to their owning companies because the same company usually has the same security reporting channel for all its products.
To avoid overwhelming the app companies, we first reported app secret leaks of one app for each company and continued to report secret leaks of other apps of the same company if we received feedback.
We also reported an issue identified in our characteristic study.

App developers have different perspectives on their leaked app secrets.
We found that two exploitable credentials in two apps had been disabled by their developers when we planned to report the leaks. 
There are nine apps whose issue reporting pages suggest that there is no need to report Google Maps API key leaks or DoS attacks. 
As a result, we did not submit reports to the developers of these apps.
For the other apps, we submitted 16 issue reports in total.
We received different responses from the app developers:
\begin{itemize}[leftmargin=*]
    \item Category-1: Three apps confirmed our reported secret leaks and repaired them.
    All the leaked secrets of these three apps were identified in more than one version of their host apps in our dataset.
    A secret extracted in one of these three apps has been used since 2019 and was found in six different apps in dataset~2021. This may suggest that the leaked secret has already been abused by some apps.
    One CVE was also filed for one of our reported secret leak~\cite{mycve}.
    \item Category-2: Three apps indicated that they would tolerate the leak of the secrets at the app end.
    One app explained that the company would monitor the credential usages for the time being.
    Should fraudulent usages be found at scale, actions will be taken to block the requests and change the credentials.
    One app indicated that it can get refunds for fraudulent uses of the leaked credentials from the concerned service provider, so the company decided to tolerate the leak.
    The other app acknowledged that monetary losses can be induced by the leaked credentials, but an economical way to perfectly protect the credentials has not yet been found. 
    As a result, the app would tolerate the leak at this stage. 
    \item Category-3: Five apps informed us that the issues were reported earlier by others, but they could not share information about the original reports due to their disclosure policies.
    \item Category-4: For the remaining reports, the developers are still analyzing the issues. We have not received their responses yet.
\end{itemize}

We can observe from the responses that app secret leaks can have two different impacts: 
First, the leaks are concerning. They could be present in the apps for a long while without being noticed. This applies to Category-1.
Second, the leaks can be tolerated. This applies to Category-2. 
The apps either monitor the fraudulent uses of their secrets or considered that the monetary losses induced by the leaked secrets are tolerable.
However, it is non-trivial for us as an outsider to judge whether a leaked secret is concerning or tolerable.
In future, we will explore how to assess the security impact of each secret leak issue to help categorize the leaked secrets.
We will also gradually release the details of these issues after obtaining disclosure permissions from the app developers.


    From our discussions with developers, we found that the primary barrier to properly protecting app secrets is the implementation overhead of secure storage.
    One developer indicated that they are reluctant to fix the issue because implementing the fix would require a redesign of the app.
    Similar findings were found in Stack Overflow posts~\cite{basak2023challenges,stackoverflowstorage}.
    Developers are struggling to find ways to securely store app secrets.
    This highlights a need for future research to develop tools that assist developers in refactoring their code to enhance the security of app secret storage.

\section{Discussions}
\subsection{Threats to Validity}
In our characteristic study, we make conclusions based on a sample set of 575 secret candidates, which may not represent all the cases of Android app secret leakage.
The sampling allows us to have sufficient time to inspect each sampled case carefully.
Our adopted sample size (575) has a confidence level of 95\% with a 4.09\% margin of error for an unlimited population size~\cite{samplesize}.
This exceeds the commonly adopted confidence level and margin of error of the sample sizes adopted by other empirical studies in the computer science discipline~\cite{harty2021logging,liu2021characterizing}.

The adopted approach using regular expressions to extract app secrets can have false negatives.
Our regular expression in Listing~\ref{listing:regex} will fail if app developers did not use meaningful variable names or the apps were obfuscated.
Nonetheless, the approach has helped us obtain a dataset of a sufficient number of app secrets for manual inspection.
Our study provides initial analysis results showing that app secret leakage can be prevalent in Android apps and inspiring future studies.

    Our study focuses on popular apps from Google Play, which may not represent less popular apps.
    This poses threats to the generalizability of our results.
    We chose popular apps because they are more likely to be well-maintained. Identifying leaked secrets in these popular apps underscores the significance of the problem: even developers of well-maintained apps can neglect the importance of protecting app secrets. This highlights the need to raise awareness of the issue among all Android developers.

Our manual inspection may be subject to errors.
To mitigate this problem, we followed the open-coding procedure~\cite{holton2007coding}.
All the secret candidates were cross-checked by two co-authors.
The two co-authors also thoroughly discussed each secret candidate and finally reached a consensus.

\subsection{Limitations and Future Work}\label{ssec:limitations}
We used regular expressions to detect leaked app secrets (Listing~\ref{listing:regex} and Table~\ref{tab:credential_detection}).
However, this technique is not robust.
They can fail if the app secrets are constructed via string computations.
They can also fail if the app or the format of the secrets are updated.
In the future, we plan to improve the app secret detection techniques by combining program analysis and machine learning techniques.
Program analysis techniques can recover string values used in apps and machine learning techniques can determine whether a string is an app secret.
We will also explore using large language models to identify app secrets.
We conducted a preliminary study to ask ChatGPT to identify secrets in Android apps.
We inputted 50 of our manually-labeled strings to ChatGPT, among which, 30 are manually identified as app secrets.
ChatGPT reported 48 of them are potential sensitive secrets, achieving a precision of 62.5\%.
The result shows that directly using the strings as inputs for large-language models cannot precisely identify app secrets.
This can pertain to the fact that a lot of random strings are not app secrets (e.g., an identifier to represent a preference item).
In the future, we plan to integrate more context information with the extracted strings to help large language models effectively detect app secrets.

\revision{
In addition, we plan to extend our empirical study to evaluate the performance of different app secret detection tools and conduct another empirical study based on the latest Android apps.
Inspired by Chen et al.~\cite{chen2020empirical}, we plan to first collect a set of state-of-the-art secret detection tools and evaluate their performance in terms of precision and recall.
We will then leverage the best-performing tools to analyze the current state of the leaked secrets in Android apps.
To this end, we plan to update our dataset by re-collecting a large dataset from Google Play or using well-maintained datasets such as AndroZoo~\cite{allix2016androzoo}.
}

Our characteristic study shows that there are app credentials for back-end cloud services that are private to the apps.
In this paper, we focused on exploiting app credentials of commonly-used third-party cloud services.
In the future, we will also study the exploitation of app credentials of app-specific back-end services and propose techniques to detect their leakage.

App developers may consider some app secret leaks tolerable.
It is difficult for us to distinguish whether the leakage of a secret is tolerable or not.
In the future, we plan to explore how to estimate the impact of the secret leaks.
Such information can help assess whether a secret leak issue is tolerable or not.
\subsection{Ethical Consideration}
We take ethical considerations carefully in our study.
First, we studied APK files that are publicly available on Google Play.
We identified different types of app secret information and did not disclose any of them.
Second, when studying and validating the extracted secrets, we never attacked the apps.
For the extracted secrets for cloud services, we validated them with minimal test cases that will not alter or expose any sensitive data.
For example, for Facebook and Twitter keys and secrets, we consider them as valid if they can be used to generate a token.
We did not further leverage the token to further access any sensitive information.
For the encryption keys, we did not use them to decode any other sensitive information in the apps.
Third, when reporting the leaks to developers, we only include the identified secrets in private bug reports that are not publicly accessible.
Finally, in this paper, we do not disclose the complete information of the apps and our extracted secrets.
In stead, we process the values and present partial information.
Due to ethical considerations, we cannot share our dataset and labeling results to the public. However, we are willing to share the dataset with verified benign researchers upon request.
\section{Related Work}
Meli et al.~\cite{meli2019bad} characterized secret leakage in GitHub repositories.
In their work, they also used regular expression to extract Cloud service credentials of certain services but did not examine whether the leaked credentials are exploitable.
In comparison, our work focus on app secret leakage in Android apps.
We conducted the first systematic study to characterize app secret leakage issues in Android apps by manually analyzing leaked app secrets and detecting exploitable app secrets in popular Android apps.
In the following, we mainly discuss related work in the context of Android apps.

\textbf{Android User Private Information Leakage.}
A large body of studies has been proposed to protect user private information in Android apps from being leaked.
For example, various static and dynamic techniques were proposed to detect privacy leaks by analyzing information flows in Android apps~\cite{enck2014taintdroid,arzt2014flowdroid}.
Gamba et al.~\cite{gamba2020analysis} analyzed pre-installed apps in customized Android OSes and showed that the pre-installed apps can also leak users' private data.
Other work investigated the problems of the permission model in Android systems~\cite{shen2021can}, or how apps can circumvent the permission systems and steal user private information~\cite{reardon201950,wijesekera2015android}.
Our study has a different focus from these works.
While they studied user private information leakage, we focus on app secret leakage in Android app code.
App secrets are sensitive credentials that belong to Android apps and are bundled in app code.
Leakage of app secrets is a different problem from that of user private information and thus, should be detected and mitigated differently.

Zuo et al.~\cite{zuo2019does} investigated user private data leakage induced by misuses of Cloud Service API credentials in Android apps, which is the most relevant to our work.
They only focused on detected key misuses of three back-end as a service (BaaS) cloud services including Azure Storage, Azure Notification Hub, and AWS.
In comparison, we conducted a characteristic study to draw a bigger picture of app secret leakage issues in Android.
We disclosed that app secret leaks are not only restricted to cloud service credentials (Table~\ref{tab:RQ1Category}), and they can induce various consequences other than user private data leakage (Section~\ref{ssec:RQ2}).

\textbf{OAuth and Mobile Apps.}
Studies were conducted to analyze security problems on using OAuth in mobile apps.
Chen et al.~\cite{chen2014oauth} analyzed Android apps using OAuth and derived common implementation issues.
They conducted a field study and disclosed 59.7\% of their studied apps were not correctly implementing the protocol.
Rahat et al.~\cite{al2019oauthlint} proposed \textsc{OAuthLint} to detect OAuth implementation problems in Android apps via static analysis.
Our work has a different scope from these works.
They focus on exposing different kinds of security issues in OAuth implementation in Android apps but our work focuses on exposing different kinds of app secrets that are leaked in Android apps: OAuth credentials are just one kind of app secrets.
With regard to OAuth credential leakage in Android apps, both of these papers reported hard-coded OAuth credentials in Android apps.
However, they simply discussed the existence of such cases and did not propose techniques to detect leaked OAuth credentials, discuss exploitation approaches, or assess their prevalence.
Viennot et al.~\cite{viennot2014measurement} also reported cases of leaked AWS API keys and OAuth credentials in their analysis of Google Play apps.
However, they merely reported the statistics of leaked credentials of certain cloud services and did not assess whether the credentials are exploitable.
Our work differs from theirs in that
we demonstrated the prevalence of exploitable app secrets even among the most popular apps on Google Play.
Our study highlights that Android app developers continue to struggle with the problem of hard-coded secrets despite the well-known risks.
Moreover, Observation 6 shows that app secrets can remain exposed in apps for long periods. This emphasizes the critical need to assist app developers in effectively securing their secrets.

    A blog from a commercial organization also reports the presence of hard-coded secrets in Android apps~\cite{zdnet}.
    However, it merely highlights the issue without providing further analysis on how the secrets were stored.
    In contrast, we conducted an in-depth analysis of leaked app secrets in Android apps.
    For example, our observations on the storage of app secrets can guide the design of improved detection tools.
    Additionally, our analysis of the life span of these secrets shows that developers often neglect this issue for extended periods.
    Our analysis demonstrates that the problem of hard-coded secrets is more severe than indicated by commercial reports.

\section{Conclusion}
In this paper, we conducted the first systematic study to characterize app secret leakage issues in Android apps.
We categorized the leaked app secrets according to their usages, assessed their security impact, and summarized the bad practices in storing app secrets.
Based on our findings, we discussed the possible countermeasures to alleviate app secret leakage issues and derived techniques to detect leaked app secrets of the most popular cloud services.
Our study demonstrated that app secrets are not far from being stolen: they can be easily extracted with simple techniques even in the most popular Android apps on Google Play.

In the future, we plan to further improve the robustness of our app secret detection approaches to detect more kinds of app secrets and to expand the scale of our study.
We will also perform in-depth analysis on more kinds of app secrets to better assess the security impact of their leakage.

\textbf{Data availability statement.} We cannot make the dataset and study results publicly available since the results involve app sensitive information. The identified leaked app secrets may lead to attacks to the studied apps. However, we are willing to share the dataset to verified researchers upon request.
\section*{Acknowledgements}
We would like to thank all the reviewers for their constructive feedback that helped us improve this paper.
This work was supported by Natural Sciences and Engineering Research Council of
Canada Discovery Grant (Grant No. RGPIN-2022-03744 and Grant No.
DGECR-2022-00378), FRQNT/NSERC NOVA program (Grant no. 2024-NOVA-346499), The Hong Kong Postdoctoral Fellowship (Grant no. PDFS2021-6S06), Hong Kong Research Grant Council/General Research Fund (Grant No. 16205821), and that start-up grant of City University of Hong Kong (Grant no. 9610676).

\section*{Conflict of Interest}
The authors declared that they have no conflict of interest.


\newpage
\bibliographystyle{IEEEtran}
\balance
\Urlmuskip=0mu plus 1mu
\bibliography{bibli} 
\appendix
In Section~\ref{ssec:RQ1}, we found credentials of the following cloud services:
\begin{enumerate}
    \item Adjust~\cite{adjust}
    \item Adobe IMS API~\cite{adobe}
    \item AirShip~\cite{airship}
    \item Algolia~\cite{Algolia}
    \item Alipay~\cite{alipay}
    \item Amplitude~\cite{amplitude}
    \item AppDynamics~\cite{appdynamics}
    \item AppMetrica~\cite{appmetrica}
    \item Appodeal~\cite{appodeal}
    \item AppsFlyer~\cite{Appsflyer}
    \item AWS S3~\cite{aws}
    \item Azure Speech Service~\cite{azurespeech}
    \item BitMovin~\cite{bitmovin}
    \item Box Platform~\cite{box}
    \item BuzzAd~\cite{buzzad}
    \item CodePush~\cite{codepush}
    \item Cloudinary~\cite{cloudinary}
    \item Countly~\cite{countly}
    \item Dropbox~\cite{dropbox}
    \item Facebook~\cite{facebook}
    \item Flurry~\cite{flurrymain}
    \item Firebase Cloud Messaging~\cite{fcm}
    \item Google Analytics~\cite{googleanalytics}
    \item Google Cloud Vision~\cite{vision}
    \item Google Maps Platform~\cite{googlemapsapi}
    \item Google Translation~\cite{googletranslation}
    \item HockeyApp~\cite{hockeyapp}
    \item Iterable~\cite{iterable}
    \item Keen Analytics~\cite{keen}
    \item Leanplum~\cite{leanplum}
    \item Line~\cite{line}
    \item MapBox~\cite{mapbox}
    \item Midtrans~\cite{midtrans}
    \item Mintegral~\cite{mintegral}
    \item Mi Push~\cite{mipush}
    \item MixPanel~\cite{mixpanel}
    \item mParticle~\cite{mpariticle}
    \item myTracker~\cite{mytracker}
    \item New Relic~\cite{newrelic}
    \item Open Weather~\cite{openweather}
    \item Parse~\cite{parse}
    \item Plaid~\cite{plaid}
    \item reCAPTCHA~\cite{recaptcha}
    \item Salesforce~\cite{salesforce}
    \item Schoology~\cite{schoology}
    \item Sift~\cite{sift}
    \item Stripe~\cite{stripe}
    \item Tencent~\cite{qq}
    \item They Said So~\cite{theysaidso}
    \item TMap API~\cite{tmap}
    \item Twitter~\cite{twitter}
    \item WeChat~\cite{wechat}
    \item Weibo~\cite{weibo}
    \item Yammi~\cite{yammi}
    \item YouTube~\cite{youtube}
\end{enumerate}
\end{document}